\documentclass[pra,10pt,twocolumn,superscriptaddress,floatfix,showpacs]{revtex4-1}

\usepackage{subfig}
\usepackage{graphicx} 
\usepackage{epstopdf}
\usepackage[utf8]{inputenc}
\usepackage[T1]{fontenc}     
\usepackage[british]{babel}  
\usepackage[sc,osf]{mathpazo}
\usepackage[scaled=0.86]{berasans}  
\usepackage[colorlinks=true, citecolor=blue, urlcolor=blue]{hyperref}  
\usepackage[babel]{microtype}  
\usepackage{amsmath,amssymb,amsthm,bm,amsfonts,mathrsfs,bbm} 

\usepackage{xspace}  
\usepackage{pgfplots}
\usepackage{xcolor,colortbl}
\def\ba{\begin{equation}}
\def\ea{\end{equation}}
\def\bea{\begin{eqnarray}}
\def\eea{\end{eqnarray}}
\def\ben{\begin{equation*}}
\def\een{\end{equation*}}
\def\bean{\begin{eqnarray*}}
\def\eean{\end{eqnarray*}}
\def\bma{\begin{mathletters}}
\def\ema{\end{mathletters}}
\def\bi{\begin{itemize}}
\def\ei{\end{itemize}}

\newcommand{\be}{\begin{equation}}
\newcommand{\ee}{\end{equation}}

\newcommand{\kommentar}[1]{}

\newcommand{\forget}[1]{}

\begin{document}

\title{Detecting Two Qubit Both-Way Positive Discord States}

\author{Kaushiki Mukherjee}
\email{kaushiki_mukherjee@rediffmail.com}
\affiliation{Department of Mathematics, Government Girls' General Degree College, Ekbalpore, Kolkata-700023, India.}

\author{Sumana Karmakar}
\email{sumanakarmakar88@gmail.com}
\affiliation{Department of Mathematics, Heritage Institute of Technology, Anandapur, Kolkata-700107, India.}

\author{Biswajit Paul}
\email{biswajitpaul4@gmail.com}
\affiliation{Department of Mathematics, South Malda College, Malda, West Bengal, India}

\author{Debasis  Sarkar}
\email{dsarkar1x@gmail.com, dsappmath@caluniv.ac.in}
\affiliation{Department of Applied Mathematics, University of Calcutta, 92, A.P.C. Road, Kolkata-700009, India.}


\begin{abstract}
Quantum discord plays a pragmatic role in analyzing nonclassical feature of quantum correlations beyond entanglement. It is used in several information processing protocols which lacks sufficient amount of entanglement to be used as a resource. We have provided with an analytical method of detecting quantum discord of an arbitrary two qubit state. We have formulated a set of necessary and sufficient conditions for any two qubit state to be a both-way non-zero quantum discord state. As quantum discord is asymmetric in nature, we have framed the set of if and only if conditions for a two qubit state to be classical-quantum as well for it to be quantum-classical. Interestingly, not only correlation tensor but also local Bloch vector(corresponding to the classical party) plays a role for detecting the state to be a positive discord state.
\end{abstract}

\maketitle

\section{Introduction}
The departure of quantum mechanics from classical world is guaranteed via non-classical nature of quantum correlations. Quantum entanglement is considered as the key ingredient in manifesting the deviation observed in behavior of quantum systems from their classical counterparts\cite{hor review}. However recent trend of research activities has pointed out non-classicality of unentangled quantum systems\cite{separable} thereby ensuring a finer line of demarcation between classical and quantum world. In broader sense, our present topic of discussion will be contributory in the direction of analyzing non-classicality beyond quantum entanglement.  \\
Speaking of non-classicality beyond entanglement, discovery of quantum discord may be attributed to provide a breakthrough in manifesting nonclassical correlations from separable states. The idea of quantum discord\cite{QD} is basically motivated by the existing difference between two classically equivalent definitions of mutual information if the scenario involved is quantum in nature. Further illustrations regarding quantum discord will be discussed in Sec.\ref{pre}.\\ Till date, various features of quantum discord have come to the forefront via multi-faceted exploration in this direction. For instance, if a pure bipartite state is considered, discord becomes equivalent to von Neumann entropy\cite{discord entropy}. For some unentangled states discord turns out to be non zero. Besides, not only has quantum discord been computed in diverse situations\cite{separable,pati14ii,pati14iii}, thereby introducing various measures and other allied concepts, but also has been extended beyond two party and higher dimensional quantum systems\cite{alter}. \\
As has already been mentioned that not all separable states have non zero discord. This in turn opens up the area of exploring possible ways of characterizing a separable quantum state having non vanishing discord.
In this context, formulation of a set of necessary and sufficient criteria detecting positive discord of a quantum state is required for providing a simplified way of checking utility of a known quantum state in any protocol demanding non zero discord which is otherwise hard to compute due to involvement of optimization over strategy of measurements. However, till date, there exist only some sufficient detection criteria\cite{suffi,suffis}, idea of geometric discord\cite{alter} and an algorithm involving numerical optimization to detect quantum discord\cite{adesso}. We have provided with an alternate characterization of a generalized two qubit quantum system to possess non zero discord(both-way). To be specific, we have algebraically formulated a set of necessary and sufficient criteria characterized by party independence(hence both-way) restricting parameters of a two qubit quantum state to have non zero discord under the usual assumption that the classical party performs projective measurements.\\
Extensive research works reveal the scope of utilizing quantumness(of physical quantum systems) beyond entanglement in various practical tasks\cite{discord signature1,discord signature2,discord signature3,separable}. Recent development in field of quantum information theory reveals existence of various practical scenarios involving nonclassical correlations from separable states\cite{discord review}. Some of such practical tasks are distributed algorithms and restricted quantum gates\cite{restricted gates}, deterministic quantum computation with one qubit(DQC1 model)\cite{separable}, quantum metrology protocols involving noisy states\cite{metrology noisy}, quantum enhancement protocol\cite{quantum enhancement}, multi-faceted application in many body physics such as tasks related to quantum  phase transitions\cite{qpt1,qpt2}, study of temperature, etc. However there also exist some practical scenarios such as local broadcasting\cite{local broadcasting}, state merging\cite{state merging1,state merging2}, etc where classical correlations turn out to be more efficient which in turn points out the fact that entanglement though necessary(for some advantage), may not be sufficient\cite{notsuf}. In these experimental scenarios, quantum discord plays a significant role. Besides, it is also useful in manifesting role of quantum correlations in various phenomena such as superselection\cite{superselection1,superselection2}, Maxwell's demons\cite{demons}, dynamics related to open quantum systems such as classicalization of quantum walks\cite{quantum walk}, mutual synchronization of dissipative quantum harmonic oscillators\cite{harmonic oscillator}, etc.\\
The importance of quantum discord from both theoretical and experimental perspectives provides the basic motivation of our present work. Formulation of a set of necessary and sufficient criteria detecting positive discord of a quantum state turns out to be interesting not only for the sake of enriching the study of quantum discord but also for providing a direct way of checking utility of a known quantum state in any protocol demanding non zero discord(of state involved) which is otherwise hard to compute due involvement of optimization over strategy of measurements. \\
Rest of our discussion is organized as follows. First we briefly review the mathematical pre-requisites in Sec.\ref{pre}. In Sec.\ref{work} we formulate the necessary and sufficient criteria for non zero discord followed by discussion on practical implications of those in Sec.\ref{prac}. We finally conclude in Sec.\ref{conc} together with some discussion on potential future research directions.
 \section{Preliminaries}\label{pre}
 \subsection{Bloch Vector Representation}
Any two qubit state $\rho_{AB}$ shared between two parties($A,B$) can be represented as:
\begin{equation}\label{st4}
\small{\rho_{AB}}=\small{\frac{1}{4}(\mathbb{I}_{2\times2}+\vec{m}.\vec{\sigma}\otimes \mathbb{I}_2+\mathbb{I}_2\otimes \vec{n}.\vec{\sigma}+\sum_{i_1,i_2=1}^{3}t_{i_1i_2}\sigma_{i_1}\otimes\sigma_{i_2})}
\end{equation}
 where $\vec{\sigma}$$=$$(\sigma_1,\sigma_2,\sigma_3), $ $\sigma_{i_k}$ are the Pauli operators along three perpendicular directions($i_k$$=$$1,2,3$). $\vec{m}$$=$$(m_1,m_2,m_3)$ and $\vec{n}$$=$$(n_1,n_2,n_3)$ denote the local Bloch vectors($\vec{m},\vec{n}$$\in$$\mathbb{R}^3$) of party $A$ and $B$ respectively and $(t_{i,j})_{3\times3}$ denotes the correlation tensor $T$ which is a real matrix.
 The components $t_{i_1i_2}$ are given by $t_{i_1i_2}$$=$$\textmd{Tr}[\rho_{AB}\,\sigma_{i_1}\otimes\sigma_{i_2}].$ Considering each row $\alpha_i(i$$=$$1,2,3)$ of $T$ as a row vector $\vec{\alpha_i}$, let us define $|\vec{\alpha}_i|$$=$$\sqrt{t_{i1}^2+t_{i2}^2+t_{i3}^2}.$ Similarly considering each column $\beta_i(i$$=$$1,2,3)$ of $T$ as a column vector $\vec{\beta_i}$ $=$$(t_{1i},t_{2i},t_{3i})$, let us define $|\vec{\beta}_i|$$=$$\sqrt{t_{1i}^2+t_{2i}^2+t_{3i}^2}.$\\
 \subsection{Quantum Discord}
Ollivier and Zurek\cite{QD} introduced the concept of \textit{Quantum Discord} as a measure of genuine quantum correlation. Total correlation, i.e., the total amount of classical and quantum correlations of a bipartite state $\rho_{AB}$ is given by its quantum mutual information
 \begin{equation}\label{mutual information}
I(A:B)=\mathbb{S}(A)+\mathbb{S}(B)-\mathbb{S}(AB).
 \end{equation}
 where $\mathbb{S}(X)$$=$$-\textmd{Tr}(\rho_X \log_2\rho_X)$ is the Von Neumann entropy of a state $\rho_X$ of the system $X$.
Whereas, its classical correlation is captured by:
 \begin{equation}\label{classical correlation}
 \begin{split}
J(B/A)=\max_{\{\vec{w}^a\}}[\mathbb{S}(B)-\mathbb{S}(B|\{\Pi_j^a\})]
  \end{split}
 \end{equation}
Here, maximum is taken over all possible directions $\vec{w}^a$(unit vector) of complete set ($j$$=$$0,1$ for qubit systems) of orthogonal projections $\{\Pi_{j}^a\}_j$ on Alice's($A$) subsystem: $ \Pi_{j}^a$$=$$\frac{1}{2}(\mathbb{I}_2+(-1)^j \vec{w}^a.\vec{\sigma})$
with $j$$=$$0,1$ corresponding to two outputs of A's orthogonal projection along $\vec{w}^a.$ $\mathbb{S}(B|\{\Pi_j^a\})$$=$$\sum_j P_j\mathbb{S}(\rho_{\small{|j}}^b)$ with
$\rho_{\small{|j}}^b(j$$=$$0,1)$ denoting conditional states of Bob($B$) based on $A$ obtaining output $j$ and $P_j$ is the probability of $A$ obtaining output $j$ while projecting along $\vec{w}^a,$ $P_j$$=$$\textmd{Tr}[(\Pi_{j}^a\otimes\mathbb{I}_{\small{2}}).\rho_{AB}](j$$=$$0,1).$
\textit{Quantum discord} \cite{QD} of bipartite state $\rho_{AB}$ is defined as the difference between its total correlation and classical correlation:
 \begin{equation}\label{QD}
  \mathbb{D}(B/A)=I(A:B)-J(B/A).
 \end{equation}
In general, quantum discord is asymmetric in nature, i.e., $\mathbb{D}(B/A)\ne \mathbb{D}(A/B)$.\\
For a bipartite state $\rho_{AB}$, $\mathbb{D}(B/A)$(Eq.(\ref{QD})) is zero if and only if there exists a complete set of rank-one orthonormal projectors $\{\Pi_{j}^a\}_j$ such that:
\begin{equation}\label{cq}
    \rho_{AB}=\sum_{j=0}^1P_j \Pi_{j}^a\otimes\rho_{|j}^b
\end{equation}
These states(Eq.(\ref{cq})) are usually known as \textit{classical-quantum} states: classical with respect to subsystem A and quantum with respect to subsystem B.
Similarly, $\mathbb{D}(A/B)$ is zero if and only if $\rho_{AB}$ is a \textit{quantum-classical} state (i.e., quantum with respect to system A and classical with respect to system B):
\begin{equation}\label{qc}
    \rho_{AB}=\sum_{j=0}^1Q_j \rho_{|j}^a\otimes\Pi_{j}^b
\end{equation}
where the notations are analogously defined exchanging roles of parties $A$ and $B.$\\
For remaining part of our analysis we will refer to $\mathbb{D}(A/B)$ and  $\mathbb{D}(B/A)$ as one-way discord. $\rho_{AB}$ will be referred to as a positive discord(both-way) state if $\mathbb{D}(B/A)$ and $\mathbb{D}(A/B)$ both turn out to be positive for $\rho_{AB}.$\\
\subsection{Entanglement of Formation}
Entanglement of formation\cite{hor review} quantifies the minimal possible average entanglement over all pure state
decomposition of a bipartite state $\rho_{AB}$
\begin{equation}
 \mathbb{E}_F(A:B)=\min_{\{(p_i,|\psi_i\rangle_{AB}):\rho_{AB}=\sum_ip_i|\psi\rangle_{AB}\langle\psi|\}}\sum_ip_iE(|\psi_i\rangle_{AB})
 \end{equation}
Entanglement of formation of a two qubit mixed state $\rho_{AB}$ is defined as
 \begin{equation}\label{entf}
 \mathbb{E}_F(A:B)=H(C(\rho_{AB}))
 \end{equation}
 where $C(\rho_{AB})$$=$$\max\{0,\sqrt{\lambda_1}-\sqrt{\lambda_2}-\sqrt{\lambda_3}-\sqrt{\lambda_4}\}$, $\lambda_i$'s are the eigenvalues, in decreasing order, of the Hermitian matrix $\rho_{AB}\tilde{\rho}_{AB}$ and $\tilde{\rho}_{AB}$$=$$\sigma_2\otimes\sigma_2\rho_{AB}\sigma_2\otimes\sigma_2$. $\sigma_2$ is the Pauli matrix and $H(x)$$=$$-x\log_2x-(1-x)\log_2(1-x)$, $x\in[0,1]$.\\
\\
\subsection{X states}
This class of two qubit states\cite{eberlyqic} is given by:\\
$\chi=\small{x_1}|00\rangle \langle 00|+\small{x_2}|01\rangle\langle 01|+\small{x_3}|10\rangle\langle10|
+\small{x_4}|11\rangle\langle11|$
\begin{equation}\label{st9}
\small{y_1}|00\rangle\langle11|+\small{y_1^{*}}|11\rangle\langle00|+\small{y_2}|01\rangle\langle10|+\small{y_2^{*}}|10\rangle\langle01|,
\end{equation}
where $x_1+x_2+x_3+x_4$$=$$1$. Non-negativity demands $|y_1|^2\leq x_1x_4$ and $|y_2|^2\leq x_2x_3$.
This family of states, being well known for their utility in various experimental scenarios\cite{Rau}, includes Bell diagonal states and hence also Werner state\cite{werner89}. X states are also studied in condensed matter systems and in various other fields of quantum mechanics.
We now discuss the results in the next section.
\section{Necessary and Sufficient Criteria for both-way Positive discord}\label{work}
For our purpose we have considered the following well-known fact(exploiting relation between discord and existence of hybrid forms of classical and quantum states\cite{QD}) that for a bipartite state $\rho_{AB}$, $\mathbb{D}(B/A)$(Eq.(\ref{QD})) is zero if and only it is a classical-quantum state(Eq.(\ref{cq})).\\
\textit{Theorem.1:} Discord $\mathbb{D}(B/A)$$=$$0$ if and only if correlation tensor $T$ of $\rho_{AB}$ is either a (i) null matrix or (ii) of rank one together with the local Bloch vector $\vec{m}$ satisfying one of the following criteria:
\begin{enumerate}
  \item If $T$ has two zero rows say $\vec{\alpha}_i$$=$$\vec{\alpha}_j$$=$$\Theta$, then either $m_k$$=$$0$ or $m_i$$=$$m_j$$=$$0.$
  \item If $T$ has one zero row say $\vec{\alpha}_i$$=$$\Theta$ and $\vec{\alpha}_j,\vec{\alpha}_k$ identical up to multiplicity, i.e., $\vec{\alpha}_j$$=$$p\vec{\alpha}_k$ then either $m_j$$=$$p m_k$ or $p m_j$$=$$-m_k.$
  \item If $T$ has all three rows identical up to multiplicity, i.e., $\vec{\alpha}_1$$=$$p_{12}\vec{\alpha}_2,$ $\vec{\alpha}_1$$=$$p_{13}\vec{\alpha}_3$ then $\vec{m}$ must satisfy $p_{13} (p_{12}m_1+m_2)+p_{12} m_3$$=$$0 $ or $m_1:m_2:m_3$$=$$p_{13}p_{12}:p_{13}:p_{12},$
\end{enumerate}
where $\Theta$$=$$(0,0,0)$ and $i,j,k\in\{1,2,3\}$ together with $i\neq j\neq k.$\\
\textit{Proof:} See Appendix.I\\
Theorem.1 provides with a set of restrictions which if satisfied by state parameters, $\mathbb{D}(B/A)$ of corresponding state $\rho_{AB}$ turns out to be zero. Conversely if discord $\mathbb{D}(B/A)$ of $\rho_{AB}$ vanishes then correlation tensor($T$) and local Bloch vector($\vec{m}$) of $\rho_{AB}$ must satisfy any one of the conditions enlisted in the theorem. Now the restrictions provided by the theorem being a set of necessary and sufficient conditions for vanishing discord, violation of that set of criteria acts as if and only if condition for positive discord(one-way) of $\rho_{AB}$. Hence, \textit{$\mathbb{D}(B/A)$$>$$0$ if and only if correlation tensor $T$ of $\rho_{AB}$ is either of rank greater than one or is of rank one together with local Bloch vector $\vec{m}$ of party $A$ violating all of the conditions($(1)$-$(3)$) laid down in Theorem.$1$.} Interestingly, for $\mathbb{D}(B/A)$$>$$0$ restrictions are imposed on correlation tensor $T$ and local Bloch vector $\vec{m}$ corresponding to $1^{st}$ party($A$) whereas the other local Bloch vector $\vec{n}$ corresponding to $2^{nd}$ party($B$) remains unrestricted. $\vec{n}$ becomes restricted if $\mathbb{D}(A/B)$ is considered. In that case however local Bloch vector $\vec{m}$ becomes unrestricted. \\
\textit{Theorem.2:} Discord $\mathbb{D}(A/B)$$=$$0$ if and only if $T$ of $\rho_{AB}$ is either a (i)null matrix or (ii)is of rank one together with the local Bloch vector $\vec{n}$ satisfying one of the following criteria:
\begin{enumerate}
  \item If $T$ has two zero columns say $\vec{\beta}_i$$=$$\vec{\beta}_j$$=$$\Theta$, then either $n_k$$=$$0$ or $n_i$$=$$n_j$$=$$0.$
  \item If $T$ has one zero column say $\vec{\beta}_i$$=$$\Theta$ and $\vec{\beta}_j,\vec{\beta}_k$ identical up to multiplicity, i.e., $\vec{\beta}_j$$=$$s\vec{\beta}_k$ then either $n_j$$=$$s n_k$ or $s n_j$$=$$-n_k.$
  \item If $T$ has all three columns identical up to multiplicity, i.e., $\vec{\beta}_1$$=$$s_{13}\vec{\beta}_2,$ $\vec{\beta}_1$$=$$s_{13}\vec{\beta}_3$ then $\vec{n}$ must satisfy $s_{13} (s_{12} n_1+n_2)+s_{12} n_3$$=$$0$ or $n_1:n_2:n_3$$=$$s_{13}s_{12}:s_{13}:s_{12},$ with $i,j,k\in\{1,2,3\}$ and $i\neq j\neq k.$
\end{enumerate}
\textit{Proof:} Similar to that of technique followed in Theorem.$1$ with rows of correlation tensor $T$  now replaced by columns and $\vec{m}$ replaced by $\vec{n}.$\\
Arguing similarly as before, it is observed that \textit{$\mathbb{D}(A/B)$$>$$0$ if and only if correlation tensor $T$ of $\rho_{AB}$ is either of rank greater than one or is of rank one together with local Bloch vector $\vec{n}$ of party $B$ violating all of the conditions($(1)$-$(3)$) prescribed in Theorem.$2.$} \\
 Combination of these two theorems directly gives the \textit{necessary and sufficient criteria for both-way positive discord:}\\
\textit{Theorem.3:} An arbitrary two qubit state $\rho_{AB}$ possesses both-way positive quantum discord if and only if either (i)correlation tensor $T$ of $\rho_{AB}$ is a matrix of rank greater than one or (ii)is of rank one along with local Bloch vectors $\vec{m}$ and $\vec{n}$ violating criteria specified by Theorem.$1$ and Theorem.$2$ respectively.\\
Clearly Theorem.$3$ gives the set of if and only if conditions for a two qubit state to have both $\mathbb{D}(A/B)$$>$$0$ and $\mathbb{D}(B/A)$$>$$0,$ i.e., to have positive discord independent of parties. Hence, irrespective of which of the two parties(sharing two qubits of $\rho_{AB}$) plays the role of the auxiliary system, $\rho_{AB}$ possesses positive discord if and only if it satisfies Theorem.$3.$\\
\subsection{Illustrations}
Consider a product state:
\begin{equation}\label{prod}
    \rho_{prod}=\frac{1}{4}(\mathbb{I}_2+\vec{v}^a.\vec{\sigma})\otimes(\mathbb{I}_2+\vec{v}^b.\vec{\sigma}).
\end{equation}
Here the local Bloch vectors are $\vec{v}^a$$=$$(v^a_1,v^a_2,v^a_3)$ and $\vec{v}^b$$=$$(v^b_1,v^b_2,v^b_3)$ and correlation tensor $T$ has three non zero rows $\vec{\alpha}_i$$=$$v^a_i(v^b_1,v^b_2,v^b_3)(i$$=$$1,2,3).$ Clearly the rows are identical up to multiplicity. Here $p_{12}$$=$$\frac{v^a_1}{v^a_2}$
and $p_{13}$$=$$\frac{v^a_1}{v^a_3}.$ So $v^a_1:v^a_2:v^a_3$$=$$p_{13}p_{12}:p_{13}:p_{12}. $ Hence $\vec{v}^a$ satisfies Theorem.$1$(criterion given by (ii.(3)) and so $\mathbb{D}(B/A)$$=$$0$. Similarly it can be checked that $\vec{v}^b$ satisfies Theorem.$2$ and hence $\mathbb{D}(A/B)$$=$$0$. This in turn points out that two qubit product state is a both-way discord zero state.\\
Next we consider $X$ states(Eq.(\ref{st9})) with real entries. For this class of states, correlation tensor is of the form:
\begin{equation}\label{x11}
    T=\textmd{diag}(2(y_1+y_2),2(-y_1+y_2),x_1-x_2-x_3+x_4),
\end{equation}
and local Bloch vectors are given by:
\begin{equation}\label{x12}
    \vec{m}=(0,0,x_1+x_2-x_3-x_4),\,\,\vec{n}=(0,0,x_1-x_2+x_3-x_4)
\end{equation}
Clearly, correlation tensor($T$) of $\chi$ is of rank $1$ if and only if any one of the following holds:
\begin{description}
  \item[a] $y_1$$=$$y_2$$=$$0$
  \item[b] $|y_1|$$=$$|y_2|$ and $x_1+x_4$$=$$x_2+x_3.$
\end{description}
Now for $\chi$ to be a classical-quantum state further restrictions are imposed on state parameters via the restricted form of local Bloch vector($\vec{m}$) of party Alice. Considering all the restrictions simultaneously(prescribed by Theorem.$1$), $\chi$ becomes a one-way discord non zero state($\mathbb{D}(B/A)$$>$$0$) if and only if each of the followings is violated:
\begin{description}
  \item[1] $y_1$$=$$y_2$$=$$0$
  \item[2] $|y_1|$$=$$|y_2|$ and $\vec{X}_1$$=$$\vec{X}_2$ where $\vec{X}_1$$=$$(x_1,x_2),$ $\vec{X}_2$$=$$(x_3,x_4).$
\end{description}
Analogously, $\mathbb{D}(A/B)$$>$$0$ if and only if $\chi$ violates both:
\begin{description}
  \item[1] $y_1$$=$$y_2$$=$$0$ and
  \item[2] $|y_1|$$=$$|y_2|$ and $\vec{Y}_1$$=$$\vec{Y}_2$ where $\vec{Y}_1$$=$$(x_1,x_3),$ $\vec{Y}_2$$=$$(x_2,x_4).$
\end{description}
Combination of above two results indicate that $\chi$ is a both-way discord non zero state if and only if both of the followings are violated:
\begin{description}
  \item[1] $y_1$$=$$y_2$$=$$0$
  \item[2] $|y_1|$$=$$|y_2|,$ $\vec{X}_1$$=$$\vec{X}_2$ where $\vec{X}_1$$=$$(x_1,x_2),$ $\vec{X}_2$$=$$(x_3,x_4)$ and $\vec{Y}_1$$=$$\vec{Y}_2$ where $\vec{Y}_1$$=$$(x_1,x_3),$ $\vec{Y}_2$$=$$(x_2,x_4).$
\end{description}
\subsection{Comparison with existing criteria}
As already mentioned before, in \cite{alter}, necessary and sufficient condition for positive quantum discord $\mathbb{D}(B/A)$ was derived for two qudit state where, using a geometrical approach, the authors designed a closed form of $\mathbb{D}(B/A)$(see Eq.(16) of \cite{alter}) restricting correlation tensor and local Bloch vector corresponding to first party. As the detection criterion provided therein depend on geometric measure, it is not considered equivalent to quantum discord(square root of twice geometric discord provides an upper bound of quantum discord for systems of two qubits\cite{adesso}). In \cite{adesso}, the authors gave a closed form of quantum discord(see Eq.(33) of \cite{adesso}) involving two measurement angles(to be minimized). As the detection algorithm involves numerical optimization\cite{adesso}, so cannot be considered as an analytical detection criterion. In contrast, the set of necessary and sufficient criteria(Theorem.$3$) proposed here is analytic. One may note that a quick observation of the theorems discussed provides with explicit forms of classical-quantum(Theorem.$1$) and also quantum-classical(Theorem.$2$) states(see Tables.II,III in Appendix). So for a problem requiring knowledge of all possible classical-quantum or quantum-classical states, our approach and hence the criteria prescribed herein turn out to be more useful. Also as Theorem.$3$ gives the necessary and sufficient criteria for a given state to be both-way non zero quantum discord state, so if a state satisfies criteria set by Theorem.$3$ then definitely the two classically equivalent definitions of mutual information are different irrespective of whether first or second party plays the role of the measurement apparatus.
\section{Practical Implications}\label{prac}
Importance of quantum discord in experimental scenarios has already been pointed out. In this context, we consider two particular protocols.\\
\textit{Quantum State Merging:} It is an information theoretic task where two parties $A$ and $B$ share a mixed quantum state aiming to merge their parts of the state  on one party's(say,$B$'s) side. The task is to be performed in such way that the purification of their shared state remains intact. For that, $A$ and $B$ need to have access to additional singlets. It was shown in \cite{state merging1,state merging2} the singlet rate required for this is the conditional entropy $\mathbb{S}(A/B)$$=$$\mathbb{S}(AB)-\mathbb{S}(B)$. Interestingly both signatures of  conditional entropy admit operational interpretation. If $\mathbb{S}(A/B)$$>$$0$, state merging is possible with this singlet rate and task will not be completed if singlet rate is less than $\mathbb{S}(A/B)$. However, if $\mathbb{S}(A/B)$$<$$0$, not only can $A$ and $B$ merge their parts of the state on $B$'s side without any additional singlet rate by using only LOCC, but also they gain an additional singlet at rate $\mathbb{S}(A/B)$ that can be used for future communication. Clearly if $\mathbb{S}(A/B)$$=$$0$ state merging is possible via LOCC only but the parties do not gain any $e$-bit. In \cite{discord int} Cavalcanti \textit{et al.} have shown that
\begin{equation}\label{sm1}
    \mathbb{D}(A/C)=\mathbb{E}_F(A:B)+\mathbb{S}(A/B)
\end{equation}
where the system $C$ purifies the state of $A$ and $B$.\\
Notably Eq.(\ref{sm1}) restricts discord in a bipartite subsystem($A,C$) with that of entanglement of formation\cite{hor review} for subsystem ($A,B$) and quantum conditional entropy for the bipartite subsystem $(A,B)$. Clearly, if a tripartite pure state $|\psi_{ABC}\rangle$ used in the protocol be such that $\mathbb{D}(A/C)$ vanishes and $\mathbb{E}_F(A:B)$$>$$0$ then $\mathbb{S}(A/B)$ is negative which in turn points out parties $A$ and $B$ can perform state merging via LOCC with both sender($A$) and receiver($B$) gaining  $\mathbb{S}(A/B)$ amount of potential that can be used for future communication. Let GGHZ state be used in the protocol:
\begin{equation}\label{ghz1}
    |\psi\rangle_{GHZ}=\cos(\zeta)|000\rangle+\sin(\zeta)|111\rangle,\,\, \zeta\in[0,\frac{\pi}{4}].
\end{equation}
For $|\psi\rangle_{GHZ}$ each of the bipartite subsystems is separable. Hence entanglement of formation in any of the subsystem is zero. Using Theorem.$3,$ it can be checked that discord in each of the possible subsystems also vanishes. This in turn implies that quantum conditional entropy vanishes in each of the bipartite subsystems(Eq.(\ref{sm1})). Hence, in this case, for every possible distribution of qubits of $ |\psi\rangle_{GHZ}$ between parties, the sender party need not communicate any information to the receiver party for state merging but no $e-$bit is gained for future use. However $e$-bit is gained if W state is considered:\\
$ |\psi\rangle_{W}= \cos(\zeta_1)|001\rangle+\sin(\zeta_1)\sin(\zeta_2)|010\rangle+$\\
\begin{equation}\label{ws1}
   \sin(\zeta_1)\cos(\zeta_2)|100\rangle,\,\, \zeta_1,\zeta_2\in[0,\frac{\pi}{2}].
\end{equation}
There exist states from W family of states, such that when used in the protocol, $\mathbb{S}(A/B)$ becomes negative implying that parties $A$ and $B$ can perform state merging with $-\mathbb{S}(A/B)$ $e-$bits left over. For instance consider the state from W family(Eq.(\ref{ws1})) for $\zeta_1$$=$$\frac{\pi}{2},$ $\zeta_2$$=$$\frac{\pi}{4}.$ Correlation tensor of the bipartite state($\rho_{AC}$,say) shared between $A$ and $C$ is a null matrix and hence $\mathbb{D}(A/C)$$=$$0.$ $\mathbb{E}_F(A:B)$$=$$1 $. So this state, if shared between $A,$ $B$ and $C,$ $A$ and $B$ perform state merging with LOCC with $1$ $e$-bit left over.
Now both GGHZ and W are genuine tripartite entangled states. If a separable tripartite state is used instead, then each of the bipartite subsystems being a product state, is both-way discord zero state. Hence here for arbitrary arrangements of qubits, there need not be any communication between sender and receiver parties for the purpose of state merging but nothing will be left for future use.\\
However this is not the case if a biseparable state with real parameters is used:
\begin{equation}\label{bis1}
     |\psi\rangle_{bis}= (s_{11}|00\rangle+s_{12}|11\rangle)\otimes(f_1|0\rangle+f_2|1\rangle), \,
\end{equation}
where $s_{11},s_{12}$ are the Schmidt coefficients and $ |s_{11}|^2+|s_{12}|^2$$=$$1,\,\,|f_{1}|^2+|f_{2}|^2$$=$$1$. Clearly $\mathbb{E}_F(A:B)$$=$$1.$ Also correlation tensor and local Bloch vector of party $A$ for the reduced bipartite state($\rho_{AC}$,say) satisfy a criterion specified in Theorem.$1.$ So if this state is used then state merging is possible for subsystem involving parties $A$ and $B$ via LOCC only and $1$  $e$-bit remains unused.\\
$\textit{Unilocal Broadcasting:}\,\,\textmd{Quantum }$broadcasting\cite{broadcasting} \textmd{is a\, generalization\, of\, quantum\, }cloning\cite{cloning}. This task mainly deals with copying a set of density operators using linear operations(unlike unitary operations for quantum cloning). Piani \textit{et al.}\cite{Piani} considered a more generalized scenario where a multipartite state($\rho$,say) is shared between three parties $A,$ $B$ and $C.$ The main task in this scenario, referred to as $local\, broadcasting$ is to broadcast $\rho$ having access to only local operations(communication between the parties is not allowed). Recently, Luo has introduced the task of $unilocal\, broadcasting$\cite{unilocal broadcasting1,unilocal broadcasting2,unilocal broadcasting3}. This scenario lies in between quantum and local broadcasting. It has been shown that in this task any classical-quantum(classical with respect to the broadcasting, i..e, sender party) bipartite state is useful and viceversa. Hence, when first party($A$) is the sender(broadcasting party), any two qubit state $\rho_{AB}$ can be used in the protocol if and only if corresponding state parameters are restricted by Theorem.$1.$ On the other hand if $A$ is at the receiving end(i.e., party $B$ broadcasts) then $\rho_{AB}$ is useful if and only if parameters of $\rho_{AB}$ abide by restrictions prescribed in Theorem.$2.$\\
As has been already pointed out before that the if and only if criteria provided in the present work have multi-faceted utility for practical purposes involving quantum states. We have included two such instances. To be precise, criteria formulated herein may be helpful to detect utility of any known two qubit bipartite state(avoiding any further measurement) in an information processing task that relies upon discord of state(involved in corresponding protocol).\\
\section{Conclusion}\label{conc}
Undoubtedly quantum discord plays a pragmatic role in exploiting quantumness from unentangled states. However, till date, given an arbitrary two qubit state, it is not possible analytically to detect whether it is a non-zero quantum discord state or not. In this letter we have provided an analytic solution to this problem by formulating a set of criteria(Theorem.$3$) detecting non vanishing quantum discord characterized by party symmetry. So given a state, one can now check whether it is discord non zero or not without subjecting it to any further measurement. \\
Traditionally, the classical party involved is assumed to perform only projective measurements for detecting or for measuring quantum discord. Recently more generalized measurements(POVMs)\cite{pm1,pm2} are considered. Though we have followed the traditional path only, but our approach(being completely algebraic) can be easily generalized so as to derive the necessary and sufficient conditions while considering POVMs.

\section{Appendix}
\textit{Proof of Theorem.1}
Consider the Bloch sphere representation(Eq.(\ref{st4})) of an arbitrary two qubit state $\rho_{AB}.$ By applying suitable local unitary operations, state $\rho_{AB}^{'}$ can be obtained such that:
\begin{equation}\label{a1}
    \rho_{AB}^{'}=\small{\frac{1}{2^2}(\mathbb{I}_{2\times2}+\vec{a}.\vec{\sigma}\otimes \mathbb{I}_2+\mathbb{I}_2\otimes \vec{b}.\vec{\sigma}+\sum_{i=1}^{3}r_{ii}\sigma_{i}\otimes\sigma_{i})},
\end{equation}
where correlation tensor matrix($T$) has been reduced to a diagonal matrix $R$$=$$\textmd{diag}(r_{11},r_{22},r_{33}).$ The elements $r_{ii}(i$$=$$1,2,3)$ correspond to three eigen values of real symmetric matrix $T^{t}T.$ As discord remains invariant under local unitary operations, hence if $\mathbb{D}(B/A)$$=$$0$ for $\rho_{AB}^{'}$, then $\mathbb{D}(B/A)$ vanishes for $\rho_{AB}$ also. So we first consider $\rho_{AB}^{'}$ and simplify the criteria(for vanishing discord) satisfied by parameters of $\rho_{AB}^{'}$ and then derive the restrictions which in turn will be imposed on parameters of the original two qubit state $\rho_{AB}$.\\
Now, as already mentioned in the main text, $\mathbb{D}(B/A)$ vanishes if and only if $\rho_{AB}^{'}$ is a classical-quantum state(Eq.(\ref{cq})). We will search all such $\rho_{AB}^{'}$ for which there exists $\vec{w}^a$ so that $\rho_{AB}^{'}$ can be written in form given by Eq.(\ref{cq}), i.e., $\rho_{AB}^{'}$ is a classical-quantum state. Hence we assume existence of $\vec{w}^a$ and explore the restrictions imposed on state parameters of $\rho_{AB}^{'}$ therein. For that, we compare the coefficients of $|ij\rangle\langle kl|(i,j,k,l\in\{0,1\})$ coming from Bloch-sphere representation of $\rho_{AB}^{'}$ given by Eq.(\ref{a1}) and that coming due to $\rho_{AB}^{'}$ being considered as a classical-quantum state(eq.(\ref{cq})). Clearly, coefficient of $|ij\rangle\langle kl|$ should be same in both representations $\forall i,j,k,l\in\{0,1\}.$ Comparing coefficient of $|00\rangle\langle 00|$ and $|01\rangle\langle 01|$ we get respectively:
\begin{equation}\label{a2}
    -a_3+a_1 w^a_1w^a_3+a_2 w^a_2 w^a_3+r_{33}(-1+(w^a_3)^2)+(w^a_3)^2 a_3=0.
\end{equation}
and
\begin{equation}\label{a3}
    -a_3+a_1 w^a_1w^a_3+a_2 w^a_2 w^a_3+r_{33}(1-(w^a_3)^2)+(w^a_3)^2 a_3=0.
\end{equation}
Subtracting Eq.(\ref{a3}) from Eq.(\ref{a2}) we get:
\begin{equation}\label{a4}
    r_{33}(1-(w^a_3)^2)=0.
\end{equation}
Now comparing coefficient of $|11\rangle\langle 00|$ and equating real and imaginary parts we get respectively:
\begin{equation}\label{a5}
    -(w^a_3)^2 r_{11}+r_{22}-(w^a_2)^2(r_{11}+r_{22})=0
\end{equation}
and
\begin{equation}\label{a6}
   w^a_1 w^a_2(r_{11}+r_{22}) =0.
\end{equation}
Then comparing coefficient of $|10\rangle\langle 01|$ and equating real and imaginary parts we get respectively:
\begin{equation}\label{a7}
    -(w^a_3)^2 r_{11}-r_{22}+(w^a_2)^2(-r_{11}+r_{22})=0
\end{equation}
and
\begin{equation}\label{a8}
   w^a_1 w^a_2(r_{11}-r_{22}) =0.
\end{equation}
Subtracting Eq.(\ref{a7}) from Eq.(\ref{a5}) we get:
\begin{equation}\label{a9}
    r_{22}(1-(w^a_2)^2)=0.
\end{equation}
Adding Eq.(\ref{a7}) to  Eq.(\ref{a5}) we get:
\begin{equation}\label{a10}
    r_{11}(1-(w^a_1)^2)=0.
\end{equation}
Again $\vec{w}^a$ being a unit vector:
\begin{equation}\label{a11}
    (w^a_1)^2+(w^a_2)^2+(w^a_3)^2=1
\end{equation}
Simultaneous consideration of eqs.(\ref{a4},\ref{a9},\ref{a10},\ref{a11}) implies that atleast two of $r_{11},$ $r_{22}$  and $r_{33}$ must be $0.$\\
Hence, without any loss of generality, $\vec{w}^a$ exists if and only if either (i) all $r_{ii}$$=$$0(i$$=$$1,2,3)$ or (ii) $r_{11}$$=$$r_{22}$$=$$0$ but $r_{33}\neq 0$ and $(w^a_3)^2$$=$$1,$ i.e., $\vec{w}^a$$=$$(0,0,\pm 1).$ We consider case(i) and (ii) separately.\\
Case(i): $ r_{11}$$=$$r_{22}$$=$$r_{33}$$=$$0$:\\
Comparing coefficient of $|00\rangle\langle 00|$ we get:
\begin{equation}\label{a12}
a_1 w^a_1w^a_3+a_2 w^a_2 w^a_3+(-1+(w^a_3)^2) a_3=0.
\end{equation}
Comparing coefficient of $|01\rangle\langle 00|$ and equating real and imaginary parts we get respectively:
\begin{equation}\label{a13}
    (-1+(w^a_1)^2) a_1+a_3 w^a_1 w^a_3+a_2 w^a_1 w^a_2=0
\end{equation}
and
\begin{equation}\label{a14}
    (-1+(w^a_2)^2) a_2+a_3 w^a_2 w^a_3+a_1 w^a_1 w^a_2=0
\end{equation}
Eq.(\ref{a12})$\times w^a_2$$-$Eq.(\ref{a14})$\times w^a_3$ gives:
\begin{equation}\label{a15}
    \frac{a_2}{w^a_2}=\frac{a_3}{w^a_3}.
\end{equation}
Similarly, Eq.(\ref{a13})$\times w^a_2$$-$Eq.(\ref{a14})$\times w^a_1$ gives:
\begin{equation}\label{a16}
    \frac{a_2}{w^a_2}=\frac{a_1}{w^a_1}.
\end{equation}
Combining last two Eqs(\ref{a15},\ref{a16}), we get:
\begin{equation}\label{a17}
    \frac{w^a_1}{a_1}=    \frac{w^a_2}{a_2}=\frac{w^a_3}{a_3}=\mathbb{K}\textmd{(say)}
\end{equation}
Hence,
\begin{equation}\label{a18}
    w^a_i=\mathbb{K} a_i,\,i=1,2,3
\end{equation}
Using Eq.(\ref{a11}), we get $\mathbb{K}$$=$$\frac{1}{\sum _{i=1}^3 a_i^2}.$\\
With $r_{ii}$$=$$0(i$$=$$1,2,3),$ and $\vec{w}^a$$=$$\mathbb{K}(a_1,a_2,a_3),$ coefficient of each of the term $|ij\rangle\langle kl|(i,j,k,l\in\{0,1\})$ from Eq.(\ref{st4}) matches with that from Eq.(\ref{cq}). \\
$\therefore$ For $r_{ii}$$=$$0(i$$=$$1,2,3),$ $ \rho_{AB}^{'}$ is a classical-quantum state(Eq.(\ref{cq})) with $\vec{w}^a$$=$$\mathbb{K}(a_1,a_2,a_3)$ where $\mathbb{K}$$=$$\pm\frac{1}{\sqrt{\sum _{i=1}^3 a_i^2}}.$\\
Case(ii): Without loss of any generality, let $r_{11}$$=$$r_{22}$$=$$0$ but $r_{33}\neq 0$ and $w^a$$=$$(0,0,\pm 1).$ Following similar procedure of comparing coefficients, we get that $ \rho_{AB}^{'}$ is a classical-quantum state if $a_1$$=$$a_2$$=$$0.$\\
Discussing the two possible cases,we get in totality that $\rho_{AB}^{'}$ is a classical-quantum state if it is in one of the forms prescribed in Table.(I).
\begin{widetext}
\begin{center}
\begin{table}[htp]
\begin{center}
\begin{tabular}{|c|c|}
\hline
State&$\vec{w}^a$\\
\hline
  $\frac{1}{2^2}(\mathbb{I}_{2\times2}+\vec{a}.\vec{\sigma}\otimes \mathbb{I}_2+\mathbb{I}_2\otimes \vec{b}.\vec{\sigma})$& $\pm\frac{1}{\sqrt{\sum _{i=1}^3 a_i^2}}(a_1,a_2,a_3)$\\
\hline
$\frac{1}{2^2}(\mathbb{I}_{2\times2}+a_i\sigma_i\otimes \mathbb{I}_2+\mathbb{I}_2\otimes \vec{b}.\vec{\sigma}+r_{ii}\sigma_{i}\otimes\sigma_{i})(i$$=$$1,2,3)$& $\vec{w}^a$ has $w^a_i$$=$$\pm 1(i$$=$$1,2,3)$ and two other components are $0.$\\
\hline
\end{tabular}\\
\caption{Possible forms of $\rho_{AB}^{'}$ as a classical quantum state(Eq.(\ref{cq})) with direction along which Alice performs orthogonal measurements. }
\end{center}
  	\label{table1}
  \end{table}
  \end{center}
\end{widetext}
Now if possible, let there exist $\rho_{AB}^{'}$ such that it does not corresponds to any of the form prescribed in  Table.(I) but is a classical-quantum state. Hence neither of Case(i) nor Case(ii) holds. Now $\rho_{AB}^{'}$ is a classical-quantum state implies existence of  a suitable unit vector $\vec{w}^a$(along which Alice performs orthogonal projection). However existence of $\vec{w}^a$ demands any one of the two cases(Case(i), Case(ii)) to hold. So we arrive at a contradiction which in turn guarantees that $\rho_{AB}^{'}$ is a classical-quantum state if and only if it is one of the forms enlisted in Table.(I). \\
Having simplified the criteria to be satisfied by parameters of $\rho_{AB}^{'},$ we next analyze the restrictions imposed on parameters of $\rho_{AB}$ via the criteria satisfied by parameters of $\rho_{AB}^{'}.$ For that we consider the two cases individually.\\
Case(i): $r_{ii}$$=$$0(i$$=$$1,2,3)$\\
Here $R$ turns out to be a null matrix implying that all the eigen values of symmetric matrix $T^tT$ are $0.$ Now sum of all eigen values of $T^tT$ is given by $\sum_{i,j=1}^3 t_{ij}^2.$ Each eigen value being $0,$ we get $\sum_{i,j=1}^3 t_{ij}^2$$=$$0.$ This relation together with the fact that $T$ is a real matrix gives $t_{ij}$$=$$0\,\forall i,j\in\{1,2,3\}.$ Hence $T$ turns out to be a null matrix when $r_{11}$$=$$r_{22}$$=$$r_{33}$$=$$0.$\\
Case(ii): Any two of $r_{11},r_{22},r_{33}$$=$$0$, say $r_{jj}$$=$$r_{kk}$$=$$0$ and $r_{ii}\neq 0$ where $i\,j,k\in\{1,2,3\}$ and $i\neq j\neq k.$ This in turn implies two of the three real eigen values of $T^tT$ are $0.$ Hence, sum of product of eigen values(taking two at a time) vanishes and also the product of all three eigen values vanishes. These two conditions put restrictions over row vectors($\vec{\alpha_i}$ or simply $\vec{\alpha}_i(i=1,2,3)$) respectively as follow:
\begin{equation}\label{a19}
  \sum_{i,j=1}^3 |\vec{\alpha}_i|^2|\vec{\alpha}_j|^2= \sum_{i,j=1}^3 (\vec{\alpha}_i.\vec{\alpha}_j)^2,\,i\neq j
\end{equation}
and
\begin{equation}\label{a20}
\Pi_{i=1}^3|\vec{\alpha}_i|^2-\sum_{i,j,k=1}^3 |\vec{\alpha}_k|^2(\vec{\alpha}_i.\vec{\alpha}_j)^2+2(\vec{\alpha}_1.\vec{\alpha}_2)(\vec{\alpha}_1.\vec{\alpha}_3)(\vec{\alpha}_2.\vec{\alpha}_3),\,\small{i\neq j\neq k}
\end{equation}
Simplifying Eq.(\ref{a19}) we get:
\begin{equation}\label{a20}
\sum_{i,j=1}^3 |\vec{\alpha}_i|^2|\vec{\alpha}_j|^2\sin^2(\theta_{ij})=0,\,i\neq j
\end{equation}
where $\theta_{ij}$ denotes the angle in between row vectors $\vec{\alpha}_i$ and $\vec{\alpha}_j$($i,j\in\{1,2,3\}$ and $i\neq j$). Sum of squares of real quantities being $0,$ each of the term under summation vanishes:
\begin{equation}\label{a21}
 |\vec{\alpha}_i|^2|\vec{\alpha}_j|^2\sin^2(\theta_{ij})=0,\,i\neq j\, \textmd{and}\, i,j\in\{1,2,3\}.
\end{equation}
The possible cases which arise from Eq.(\ref{a21}) are:
\begin{description}
  \item[a] each of $\vec{\alpha}_i(i$$=$$1,2,3)$ is a zero vector $\Theta.$ In this case $T$ is a null matrix which is same as that in Case(i).
  \item[b] any two of row vectors, say $\vec{\alpha}_i$$=$$\vec{\alpha}_j$$=$$\Theta$ and $\vec{\alpha}_k\neq 0(i\neq j\neq k).$
  \item[c] any one of the row vectors say $\vec{\alpha}_i$$=$$\Theta.$ Then $\theta_{kj}$$=$$n\pi(k\neq j\neq i)$ which in turn implies that $\vec{\alpha}_j$ and $\vec{\alpha}_k$ are scalar multiple of each other, i.e., identical up to multiplicity.
  \item[d] none of $\vec{\alpha}_i$ is $\Theta$ which in turn guarantees that all three row vectors are scalar multiple of each other.
\end{description}
Combining all these possibilities we get that in Case(ii) either $T$ is a null matrix or is of rank $1.$ Hence, combining cases (i) and (ii) we can conclude that when $\rho_{AB}^{'},$ obtained from the original state $\rho_{AB}$ under application of suitable local unitary operations, is a classical-quantum state, the correlation tensor $T$ of $\rho_{AB}$ is either a null matrix or is of rank $1.$\\
Now if possible let there exists $\rho_{AB}$ with correlation tensor $T$ of rank greater than $1$ such that $\rho_{AB}^{'}$ obtained from it($\rho_{AB}$) is a classical-quantum state. But it has already been argued that  $\rho_{AB}^{'}$ is classical-quantum state if and only if the correlation tensor $R$ of $\rho_{AB}^{'}$ satisfies any of two cases(Case(i), Case(ii)) which in turn restricts correlation tensor $T$ to be either null matrix or of rank $1.$ This contradicts our assumption that $T$ is of rank greater than $1.$ Hence $\rho_{AB}^{'}$, obtained from $\rho_{AB}$ under application of suitable local unitary operations is classical-quantum state if and only if correlation tensor $T$ of original state $\rho_{AB}$ is either a null matrix or is of rank $1.$\\
Now we are assuming that there exist suitable local unitary operations which on application converts $\rho_{AB}$ to $\rho_{AB}^{'}$. This assumption in turn put further restrictions over parameters of $\rho_{AB}.$ Note from Table.I that for $\rho_{AB}^{'}$ to be a classical quantum state, apart from correlation tensor $R$, restrictions are imposed only on local Bloch vector($\vec{a}$) of party $A$ while the Bloch vector($\vec{b}$) of party $B$ remains unrestricted. Hence restrictions will be imposed on local Bloch vector $\vec{m}$ possessed by $A$ for the original state $\rho_{AB}$ while that possessed by $B$($\vec{n}$) will remain unrestricted. We next explore these restrictions.\\
\textit{Existence of local unitary operations:}
First we consider Case(i): Here correlation tensor of $\rho_{AB}^{'}$ is a null matrix which restricts correlation tensor of $\rho_{AB}$ to be a null matrix(as discussed). Hence in this case identity operations suffice to be the local unitary operations over both parties' qubits. So no further restriction is imposed over parameters of $\rho_{AB}$, i.e., both $\vec{m}$ and $\vec{n}$ remain unrestricted.\\
 Next we consider the other case, i.e, Case(ii) where the diagonal correlation tensor of $\rho_{AB}^{'}$ has only one non zero entry say $r_{kk}\neq 0$ and the local Bloch vector of party $A$ has only one non zero component($a_k\neq 0$).\\
 Before starting analysis, we point out the fact that correlation matrix $T$ being a real matrix, $T^tT,$ $TT^t,$ $R$ and all the corresponding eigen vectors are real. So henceforth we consider transpose in place of conjugate transpose. \\
Performing local unitary operations $U_A,U_B$($U_A$ on $A$'s qubit and $U_B$ on $B$'s qubit) on a state implies rotating local Bloch vectors $\vec{m}$ and $\vec{n}$(Eq.(\ref{st4})) of $A$ and $B$ respectively:\\
\begin{equation}\label{a22i}
    \vec{m}\rightarrow \vec{a}=\mathbb{Q}_A \vec{m}
\end{equation}
\begin{equation}\label{a22ii}
    \vec{n}\rightarrow \vec{b}=\mathbb{Q}_B \vec{n}
\end{equation}
\begin{equation}\label{a22iii}
    T\rightarrow R=\mathbb{Q}_A T(\mathbb{Q}_B)^{\dagger},
\end{equation}
where $\mathbb{Q}_A,\mathbb{Q}_B$ denote rotation matrices.
Now for any matrix there exists singular value decomposition(SVD). Hence for correlation tensor $T$(Eq.(\ref{st4})),
\begin{equation}\label{a22}
    T= M \Xi N,\,
\end{equation}
$\textmd{where}\,\, MM^{\dagger}$$=$$M^{\dagger}M$$=$$\mathbb{I}_3$$=$$NN^{\dagger}$$=$$N^{\dagger}N\,\, \textmd{and} \\ \Xi$$=$$\textmd{diag}(\varsigma_1,\varsigma_2,\varsigma_3).$ $M$ consists of left singular vectors of $T$(i.e. eigen vectors of $TT^t$) and $N$ consists of right singular vectors of $T$(i.e. eigen vectors of $T^tT$).
Clearly, Eq.(\ref{a22iii}) represents the singular value decomposition of correlation tensor $T:$
\begin{equation}\label{a23}
\Xi\,=\,R= \textmd{diag}(r_{11},r_{22},r_{33}).
\end{equation}
where $r_{ii}(i=1,2,3)$ denote singular values of $T$(i.e., square roots of eigen values of $T^tT$), 
\begin{equation}\label{a25}
    \mathbb{Q}_A=M^{\dagger}\, \textmd{and}\,    \mathbb{Q}_B=N^{\dagger}.
\end{equation}
Now $M$ consists of eigen vectors of $TT^t$. Denoting eigen vectors of $TT^t$ as $\vec{e}_i$$=$$(e_{i1},e_{i2},e_{i3})^t,$ we get:
\begin{equation}\label{a26}
    M^t=\left(
    \begin{array}{c}
     \vec{e}_{\small{1}}^t\\
     \,\\
      \vec{e}_{\small{2}}^t\\
      \,\\
      \vec{e}_{\small{3}}^t\\
      \end{array}
      \right)=
\left(
    \begin{array}{ccc}
     e_{11}&e_{12}&e_{13}\\
     e_{21}&e_{22}&e_{23}\\
           e_{31}&e_{32}&e_{33}\\
      \end{array}
      \right)
\end{equation}
In Case(ii), for $\rho_{AB}^{'}$ to be a classical-quantum state, local Bloch vector $\vec{a}$ of state $\rho_{AB}^{'}$ must have two zero components, say $a_i$$=$$a_j$$=$$0$. Using that fact along with the Eqs(\ref{a22i},\ref{a26}) we get:
\begin{equation}\label{a27}
    \vec{e}_i.\vec{m}=\vec{e}_j.\vec{m}=0\,\,\textmd{and}\,\, \vec{e}_k.\vec{m}=a_k,\,\, i,j,k\in\{1,2,3\},\,\,  i\neq j\neq k.
\end{equation}
If $a_k$$=$$0$ then Eq.(\ref{a27}) implies that $\vec{m}$ must be $\Theta.$ Now consider the other possibility: $a_k\neq 0$. Then $\vec{m}\neq\Theta.$ Hence $\vec{m}$ is orthogonal to both $\vec{e}_i$ and $\vec{e}_j.$ Now each of $\vec{e}_i(i$$=$$1,2,3)$ and $\vec{m}\in\mathbb{R}^3.$ Also $\vec{e}_1,\vec{e}_2,\vec{e}_3$ being eigen vectors of a $3\times 3$ real symmetric matrix($TT^t$)are orthogonal to each other. Combining all these facts it can be concluded that $\vec{m}||\vec{e}_k,$ i.e., $\vec{m}$ is an eigen vector of $TT^t.$ Now for the current case(Case(ii)), $R$ has only one non zero singular values, i.e., $TT^t$ has only one non zero eigen value. So $\vec{m}$ is either an eigen vector corresponding to eigen value $0$ or an eigen vector corresponding to the non zero eigen value. We deal with both the cases. Before that we first introduce some notations to be used henceforth.\\
\begin{itemize}
\item Let $\Gamma$$=$$TT^t$ be given by:
\begin{equation}\label{a28}
    \Gamma=\left(
    \begin{array}{c}
     \vec{\gamma}_1\\
      \vec{\gamma}_2\\
        \vec{\gamma}_3\\
      \end{array}
      \right)=\left(
    \begin{array}{ccc}
    |\vec{\alpha}_1|^2&\vec{\alpha}_1.\vec{\alpha}_2&\vec{\alpha}_1.\vec{\alpha}_3\\
   \vec{\alpha}_1.\vec{\alpha}_2&  |\vec{\alpha}_2|^2&\vec{\alpha}_2.\vec{\alpha}_3\\
      \vec{\alpha}_1.\vec{\alpha}_3& \vec{\alpha}_2.\vec{\alpha}_3&    |\vec{\alpha}_3|^2\\
      \end{array}
      \right)
\end{equation}
where $\vec{\alpha}_i(i=1,2,3)$ are the row vectors of matrix $T$(Eq.(\ref{st4})).
\item Whenever we mention $i,j,k$ we mean $i,j,k$$\in$$\{1,2,3\}$ and $ i\neq j\neq k.$
\end{itemize}
Subcase(i): We first consider the possibility that $\vec{m}$ is an eigen vector of $\Gamma$ corresponding to eigen value $0.$ Hence,
\begin{equation}\label{a29}
    \vec{\gamma}_i.\vec{m}=0\,\forall i\in\{1,2,3\}
\end{equation}
As we are dealing with Case(ii) so for $\rho_{AB}^{'}$ to be a classical-quantum state, rank of matrix $T$ must be $1.$ This in turn gives rise to three possibilities:\\
Subsubcase(i): $\vec{\alpha}_i$$=$$\vec{\alpha}_j$$=$$\Theta$ and $\vec{\alpha}_k\neq \Theta.$ Then $\Gamma$(Eq.(\ref{a28})) has only one non zero entry $|\vec{\alpha}_k|^2.$ Using this fact, from Eq.(\ref{a29}), we get that $k^{th}$ component of $\vec{m}$ must be $0.$ Hence in this case, if $\vec{\alpha}_k$ is the only non zero row vector of $T$, suitable local unitary operations exist(via which $\rho_{AB}^{'}$ obtained from $\rho_{AB}$ is classical-quantum state) if local Bloch vector possessed by $A$ for state $\rho_{AB}$ has $k^{th}$ component $0.$ \\
Subsubcase(ii): Only one row vector, say $\vec{\alpha}_i$$=$$\Theta$ whereas the other two are identical up to multiplicity: $\vec{\alpha}_j$$=$$p\vec{\alpha}_k.$ Clearly $|p|$$=$$\frac{|\vec{\alpha}_j|}{|\vec{\alpha}_k|}.$ This in turn modifies $\Gamma$ which in turn restricts components of $\vec{m}$(via Eq.(\ref{a29})) as follows:
\begin{equation}\label{a30}
    pm_j=-m_k.
\end{equation}
Hence in this case, suitable local unitary operations exist if $j^{th}$ and $k^{th}$ components of $\vec{m}$ satisfy relation given by Eq.(\ref{a30}).\\
Subsubcase(iii): All the row vectors are identical up to multiplicity: $\vec{\alpha}_1$$=$$p_{12}\vec{\alpha}_2,$ $\vec{\alpha}_1$$=$$p_{13}\vec{\alpha}_3.$ Clearly $|p_{12}|$$=$$\frac{|\vec{\alpha}_1|}{|\vec{\alpha}_2|}$ and $|p_{13}|$$=$$\frac{|\vec{\alpha}_1|}{|\vec{\alpha}_3|}.$ These in turn restrict the rows of $\Gamma.$ Consequently, Eq.(\ref{a29}) implies that the components of $\vec{m}$ should satisfy the following relation:
\begin{equation}\label{a31}
p_{13} (p_{12}m_1+m_2)+p_{12} m_3=0
\end{equation}
Subcase(ii): Next we first consider the possibility that $\vec{m}$ is an eigen vector of $\Gamma$ corresponding to the non zero eigen value(say $\epsilon$). Hence,
\begin{equation}\label{a31}
    \Gamma\vec{m}=\epsilon \vec{m}
\end{equation}
Let $\Delta$$=$$\Gamma-\epsilon \mathbb{I}_3.$ Then matrix $\Delta$ is explicitly given by:
$$ \Delta=\left(
    \begin{array}{c}
     \vec{\delta}_1\\
      \vec{\delta}_2\\
        \vec{\delta}_3\\
      \end{array}
      \right)$$
\begin{equation}\label{a32}
     =\small{\left(
    \begin{array}{ccc}
   -( |\vec{\alpha}_2|^2+|\vec{\alpha}_3|^2)&\vec{\alpha}_1.\vec{\alpha}_2&\vec{\alpha}_1.\vec{\alpha}_3\\
   \vec{\alpha}_1.\vec{\alpha}_2&   -( |\vec{\alpha}_3|^2+|\vec{\alpha}_1|^2)&\vec{\alpha}_2.\vec{\alpha}_3\\
      \vec{\alpha}_1.\vec{\alpha}_3& \vec{\alpha}_2.\vec{\alpha}_3&     -( |\vec{\alpha}_1|^2+|\vec{\alpha}_2|^2)\\
      \end{array}
      \right)}
\end{equation}
Now as $\vec{m}$ is an eigen vector of $\Gamma$ corresponding to $\epsilon,$ so:
\begin{equation}\label{a33}
  \vec{\delta}_i.\vec{m}=0\,\forall i\in\{1,2,3\}
\end{equation}
Again as in Subcase(ii), we have to consider three possibilities corresponding to three different ways of $T$ matrix(Eq.(\ref{st4})) to be of rank $1.$ :\\
Subsubcase (i):$\vec{\alpha}_i$$=$$\vec{\alpha}_j$$=$$\Theta$ and $\vec{\alpha}_k\neq \Theta.$ This in turn modifies matrix $\Delta$ and restrictions get imposed over $\vec{m}$ via Eq.(\ref{a33}): $i^{th}$ and $j^{th}$ components of $\vec{m}$ should be $0.$ \\
Subsubcase(ii): Only one row vector, say $\vec{\alpha}_i$$=$$\Theta$ whereas the other two are identical up to multiplicity: $\vec{\alpha}_j$$=$$p\vec{\alpha}_k.$ Considering the modified form of $\Delta,$ Eq.(\ref{a33}) implies that components of $\vec{m}$ must satisfy:
\begin{equation}\label{a34}
    m_j=p m_k.
\end{equation}
Subsubcase(iii): All the row vectors are identical up to multiplicity: $\vec{\alpha}_1$$=$$p_{12}\vec{\alpha}_2,$ $\vec{\alpha}_1$$=$$p_{13}\vec{\alpha}_3.$ Under this assumption Eq.(\ref{a33}) restricts the components of $\vec{m}$ as follows:
\begin{equation}\label{a35}
    m_1:m_2:m_3=p_{13}p_{12}:p_{13}:p_{12}.
\end{equation}
This ends our analysis. \\
Clearly we have considered all the possibilities for which suitable local unitary operations exist, which on application over $\rho_{AB}$, gives classical-quantum state $\rho_{AB}^{'}.$
Consideration of all the possibilities related to existence of suitable local unitary operations provide with necessary and sufficient conditions to be imposed on $\vec{m}$($\vec{n}$ may be arbitrary owing to the fact the corresponding local Bloch vector $\vec{b}$ of $\rho_{AB}^{'}$ remains unrestricted for $\rho_{AB}^{'}$ to be a classical-quantum state).
In totality, considering all possible cases, we get Theorem.$1.$ $\blacksquare$\\
\textit{Proof of Theorem.$2$}: It can be proved following a similar sort of argument as that in Theorem.$1$. To be precise, the if and only if restrictions imposed over correlation tensor $T$ of $\rho_{AB}$ and over local Bloch vector $\vec{b}$(of $\rho_{AB}^{'}$) come out in exactly the same manner as in Theorem.$1.$ For deriving the restrictions to be imposed over Bloch vector $\vec{n}$(here $\vec{m}$ remains unrestricted as the corresponding Bloch vector $\vec{a}$ of $\rho_{AB}^{'}$ remains unrestricted), we have to consider the matrix $N$ which consists of right singular vectors of $T$, i.e., eigen vectors of $T^tT.$ Consequently in place of $\Gamma$$=$$TT^t$ we need to consider $\Omega$$=$$T^tT$:
\begin{equation}\label{a36}
     \Omega=\left(
    \begin{array}{ccc}
    |\vec{\beta}_1|^2&\vec{\beta}_1.\vec{\beta}_2&\vec{\beta}_1.\vec{\beta}_3\\
   \vec{\beta}_1.\vec{\beta}_2&  |\vec{\beta}_2|^2&\vec{\beta}_2.\vec{\beta}_3\\
      \vec{\beta}_1.\vec{\beta}_3& \vec{\beta}_2.\vec{\beta}_3&    |\vec{\beta}_3|^2\\
      \end{array}
      \right)
\end{equation}
where $\vec{\beta}_i(i$$=$$1,2,3)$ denote the column vectors of $T$. Also instead of $\Delta$ we have to consider $\Lambda:$
\begin{equation}\label{a36}
      \Lambda=\small{\left(
    \begin{array}{ccc}
  -( |\vec{\beta}_2|^2+|\vec{\beta}_3|^2)&\vec{\beta}_1.\vec{\beta}_2&\vec{\beta}_1.\vec{\beta}_3\\
   \vec{\beta}_1.\vec{\beta}_2&   -( |\vec{\beta}_3|^2+|\vec{\beta}_1|^2)&\vec{\beta}_2.\vec{\beta}_3\\
      \vec{\beta}_1.\vec{\beta}_3& \vec{\beta}_2.\vec{\beta}_3&     -( |\vec{\beta}_1|^2+|\vec{\beta}_2|^2)\\
      \end{array}
      \right).}
\end{equation}
Then following the same pattern of argument we get criteria prescribed in Theorem.$2.$
\section{Appendix}
Below we enlist all possible forms of classical-quantum and quantum-classical states.
\begin{widetext}
\begin{center}
\begin{table}[htp]
\begin{center}
\begin{tabular}{|c|c|}
\hline
Row vectors of $T$& Local Bloch vector $\vec{m}$\\
\hline
  $\vec{\alpha}_i$$=$$\Theta$, $\forall\, i$& $\vec{m}$ is arbitrary\\
\hline
$\vec{\alpha}_i$$=$$\vec{\alpha}_j$$=$$\Theta$, $\vec{\alpha}_k\neq \Theta$& Either $m_k$$=$$0$ or $m_i$$=$$m_j$$=$$0$\\
\hline
$\vec{\alpha}_i$$=$$\Theta$ and $\vec{\alpha}_j$$=$$p\vec{\alpha}_k$ where $|p|$$=$$\frac{|\vec{\alpha}_j|}{|\vec{\alpha}_k|}$& Either $m_j$$=$$p m_k$ or $p m_j$$=$$-m_k$\\
\hline
$\vec{\alpha}_1$$=$$p_{12}\vec{\alpha}_2,$ $\vec{\alpha}_1$$=$$p_{13}\vec{\alpha}_3$ & $\vec{m}$ must satisfy $p_{13} (p_{12}m_1+m_2)+p_{12} m_3$$=$$0 $ \\
where $|p_{12}|$$=$$\frac{|\vec{\alpha}_1|}{|\vec{\alpha}_2|}$ and $|p_{13}|$$=$$\frac{|\vec{\alpha}_1|}{|\vec{\alpha}_3|}$&or $m_1:m_2:m_3$$=$$p_{13}p_{12}:p_{13}:p_{12}$\\
\hline
\end{tabular}\\
\caption{Possible forms of correlation tensor $T$ in terms of its row vectors $\vec{\alpha}_i$ and local Bloch vector $\vec{m}$ for which $\rho_{AB}$(Eq.(\ref{st4})) is classical-quantum state. Local Bloch vector $\vec{n}$ remains arbitrary. As already mentioned before in Appendix.I, here $i\neq j\neq k\in\{1,2,3\}.$ }
\end{center}
  	\label{table2}
  \end{table}
\begin{table}[htp]
\begin{center}
\begin{tabular}{|c|c|}
\hline
Row vectors of $T$& Local Bloch vector $\vec{n}$\\
\hline
  $\vec{\beta}_i$$=$$\Theta$, $\forall\, i$& $\vec{n}$ is arbitrary\\
\hline
$\vec{\beta}_i$$=$$\vec{\beta}_j$$=$$\Theta$, $\vec{\beta}_k\neq \Theta$& Either $n_k$$=$$0$ or $n_i$$=$$n_j$$=$$0$\\
\hline
$\vec{\beta}_i$$=$$\Theta$ and $\vec{\beta}_j$$=$$s\vec{\beta}_k$ where $|s|$$=$$\frac{|\vec{\beta}_j|}{|\vec{\beta}_k|}$& Either $n_j$$=$$s n_k$ or $s n_j$$=$$-n_k$\\
\hline
$\vec{\beta}_1$$=$$s_{12}\vec{\beta}_2,$ $\vec{\beta}_1$$=$$s_{13}\vec{\beta}_3$ & $\vec{n}$ must satisfy $s_{13} (s_{12}n_1+n_2)+s_{12} n_3$$=$$0 $ \\
where $|s_{12}|$$=$$\frac{|\vec{\beta}_1|}{|\vec{\beta}_2|}$ and $|s_{13}|$$=$$\frac{|\vec{\beta}_1|}{|\vec{\beta}_3|}$&or $n_1:n_2:n_3$$=$$s_{13}s_{12}:s_{13}:s_{12}$\\
\hline
\end{tabular}\\
\caption{Possible forms of correlation tensor $T$ in terms of its column vectors $\vec{\beta}_i$ and local Bloch vector $\vec{n}$ for which $\rho_{AB}$(Eq.(\ref{st4})) is quantum-classical state. Local Bloch vector $\vec{m}$ remains arbitrary. }
\end{center}
\label{table3}
\end{table}
\end{center}
\end{widetext}
 \end{document}